\documentclass[preprint,showpacs,showkeys]{revtex4}
\usepackage{amsmath}
\usepackage{amsmath,amssymb}

\begin{document}

\title{Spectral density method in quantum nonextensive thermostatistics and
magnetic systems with long-range interactions}

\begin{abstract}
Motived by the necessity of explicit and reliable calculations, as a valid
contribution to clarify the effectiveness and, possibly, the limits of the
Tsallis thermostatistics, we formulate the Two-Time Green Functions Method
in nonextensive quantum statistical mechanics within the optimal Lagrange
multiplier framework, focusing on the basic ingredients of the related
Spectral Density Method (SDM). Besides, to show how the SDM works, we have
performed, to the lowest order of approximation, explicit calculations of
the low-temperature properties for a quantum $d$-dimensional spin-$1/2$
Heisenberg ferromagnet with long-range interactions decaying as $1/r^{p}$ ($%
r $ is the distance between spins in the lattice).
\end{abstract}

\author{A. Cavallo}
\altaffiliation[Also at ]{Institut für Physik, Johannes Gutenberg Universität, D-55099 Mainz, Germany}
\author{F. Cosenza}
\email{cosfab@sa.infn.it}
\author{L. De Cesare}
\affiliation{Dipartimento di Fisica "E.R. Caianiello", Universita' degli Studi di Salerno,
and INFM, Unita' di Salerno, Via S. Allende, I-84081 Baronissi (SA), Italy}
\pacs{05.30.-d; 05.70.-a; 75.10.jm}
\keywords{Spectral Density Method, Nonextensive Thermostatistics}

\maketitle


\section{Introduction}

Green's functions (GF's) are currently used in many-body physics and their
power and success are widely recognized \cite{GFbooks}. There exists at
present a large variety of methods and techniques for calculation of the GF's
both in classical and quantum thermostatistics \cite
{GFbooks,Tyablikov67,GF-classical,Kalashnikov73,Campana84}. In particular,
the related \textit{spectral density method} (SDM), originally formulated by
Kalashnikov and Fradkin \cite{Kalashnikov73} for quantum many-body systems,
is a powerful nonperturbative tool which allows a direct study of the
macroscopic properties of interacting quantum and classical many-body
systems \cite{Kalashnikov73,Campana84,Campana85,CSDM04,CSDM06} also
involving phase transitions.

Recently, the increasing interest in Tsallis' nonextensive thermostatistics 
\cite{TsallisBasic-LRI}, has stimulated a lot of works \cite{GFq,CavalloNEXT}
on the extension of the GF formalism also to this Tsallis' generalized
framework of the statistical mechanics with the aim to provide new and
effective methods for dealing with realistic nonextensive problems. Along
this direction, the two-time GF technique and SDM have been formulated in
classical nonextensive thermostatistics in two our papers \cite{CavalloNEXT}
with application to the Heisenberg spin chain with short-range interactions.
Here we wish to present the extension of the same formalism in quantum
nonextensive thermostatistics by using the \textit{optimal Lagrange
multiplier} (OLM) representation \cite{MartinezFerri-OLM}, focusing on the
spectral density (SD) and its spectral decomposition for their relevance in
explicit calculations. Then, in order to show how the method works in the
nonextensive context, we apply the extended SDM to a quantum $d$-dimensional
spin-$1/2$ Heisenberg ferromagnet with long-range interactions decaying as $%
r^{-p}$ ($r$ is the spin distance and $p$ the decay exponent) and explore,
to the lowest order of approximation, the nonextensivity effects on the
low-temperature magnetic properties of the model.

In Sec. II we summarize some basic ingredients of the nonextensive quantum
thermostatistics in the OLM representation for next utility. Sections III
and IV are devoted to the extension of the GF formalism and of the SDM in
the nonextensive context, respectively. In Sec. V the extended SDM is
applied to the above mentioned spin model. Finally, some concluding remarks
are drawn in Sec. VI.

\section{The OLM representation of quantum nonextensive statistical mechanics%
}

The Tsallis' quantum thermostatistics \cite{TsallisBasic-LRI} is essentially
based on the so called $q$-entropy (with $k_{B}=\hbar =1$) 
\begin{equation}
S_{q}=\frac{1-Tr\rho _{q}^{q}}{q-1},
\end{equation}
where $\rho _{q}$ is the generalized density operator satisfying the
normalization condition $Tr\rho _{q}=1$. In this framework, the
generalization of the statistical average (referred as $q$\textit{%
-expectation value} or $q$\textit{-mean value}) for the observable $O$ is
given by 
\begin{equation}
\left\langle O\right\rangle _{q}=\frac{Tr\left[ \rho _{q}^{q}O\right] }{Tr%
\left[ \rho _{q}^{q}\right] }.  \label{Eq:qMean}
\end{equation}
Here, $q$ is a real parameter (called the \textit{nonextensivity index})
which measures the degree of nonextensivity.

Working in the canonical ensemble, the statistical operator $\rho _{q}$ can
be determined adopting the OLM constraint prescriptions in the extremization
procedure of $S_{q}$ \cite{TsallisBasic-LRI,MartinezFerri-OLM}. Then,
assuming the Hamiltonian $H$ of the system to have a complete orthonormal
set of eigenvectors $\left\{ \left| n\right\rangle \right\} $ with
eigenvalues $\left\{ \varepsilon _{n}\right\} $, the normalized probability $%
p_{n}=\left\langle n\right| \rho _{q}\left| n\right\rangle $ associated with
the $n^{th}$ eigenstate is given by 
\begin{equation}
p_{n}=Z_{q}^{-1}\left[ 1-\beta (1-q)\left( \varepsilon _{n}-U_{q}\right) %
\right] ^{\frac{1}{1-q}},  \label{Pq(Ei)}
\end{equation}
where 
\begin{equation}
Z_{q}=\sum_{n}\left[ 1-\beta (1-q)\left( \varepsilon _{n}-U_{q}\right) %
\right] ^{\frac{1}{1-q}},  \label{Zq(Ei)}
\end{equation}
and the $q$-internal energy $U_{q}=\left\langle H\right\rangle _{q}$ is
given by 
\begin{equation}
U_{q}=\overline{Z}_{q}^{-1}\sum_{n}p_{n}^{q}\varepsilon _{n}.  \label{Uq(Ei)}
\end{equation}
Here $\left\langle n\right| \rho _{q}^{q}\left| m\right\rangle =$ $%
p_{n}^{q}\delta _{nm}$ are the matrix elements of the operator $\rho
_{q}^{q} $ in the $\left\{ \left| n\right\rangle \right\} $ representation
and $\overline{Z}_{q}=Tr\left[ \rho _{q}^{q}\right] =\sum_{n}p_{n}^{q}$
denotes the pseudo-partition function in the normalized OLM framework 
($\overline{Z}_{1}$ is not merely the extensive partition function but contains
an extra factor).
In the previous equations, $\beta =1/T$ and $T$ is the thermodynamic
temperature. It is worth mentioning that, for $q<1$, the formalism imposes a
high-energy cutoff, i.e. $p_{n}=0$ whenever the argument of the power
function in Eqs. (\ref{Pq(Ei)})-(\ref{Zq(Ei)}) becomes negative \cite
{TsallisBasic-LRI}.

\section{Two-time $q$-Green functions and $q$-spectral density}

In the quantum nonextensive thermostatistics, we define the two-time \textit{%
retarded} and \textit{advanced} GF's ($q$-GF's) for two arbitrary operators $%
A$ and $B$ as 
\begin{equation}
G_{qAB}^{(\nu )}\left( t,t^{\prime }\right) =-i\theta _{\nu }\left(
t-t^{\prime }\right) \left\langle [A(t),B(t^{\prime })]_{\eta }\right\rangle
_{q},  \label{GF_qAB}
\end{equation}
where $\nu =r,a$ stands for ''retarded'' and and ''advanced'', respectively.
Here, $\theta _{a}(t-t^{\prime })=-\theta (t^{\prime }-t)$ and $\theta
_{r}(t-t^{\prime })=\theta (t-t^{\prime })$, being $\theta (x)$ the step
function; $[...,...]_{\eta }$ denotes a commutator ($\eta =-1$) or
anticommutator ($\eta =+1$); $X(t)$ (with $X=A,B$) is the usual Heisenberg
representation of operator $X$ at time $t$. Of course, for $q=1$, the
conventional formalism is reproduced. It is worth nothing that, for general
operators $A$ and $B$, one can develop the $q$-GF's framework equivalently
with commutators or anticommutators. However, for fermionic or bosonic
operators it is of course convenient to use\ in Eq. (\ref{GF_qAB}) $\eta =-1$
or $\eta =+1$, respectively.

It is now relatively simple generalize most of the basic properties of the
conventional two-time GF properties \cite{GFbooks,Tyablikov67} in
nonextensive context \cite{CavalloNEXT}.

Within equilibrium ensembles, the functions $G_{qAB}^{(\nu )}\left(
t,t^{\prime }\right) $ depend on times only through the difference $\tau
=t-t^{\prime }$ and hence one can introduce the Fourier transform 
\begin{equation}
\text{\ }G_{qAB}^{(\nu )}\left( \omega \right) =\int_{-\infty }^{+\infty
}d\tau G_{qAB}^{(\nu )}\left( \tau \right) e^{i\omega \tau }.
\end{equation}
Besides, in strict analogy with the GF extensive formalism \cite
{Tyablikov67,Kalashnikov73}, the generalized spectral density ($q$-SD) in
the $\omega $-representation is defined by 
\begin{equation}
\Lambda _{qAB}\left( \omega \right) =\int_{-\infty }^{+\infty }dte^{i\omega
\tau }\left\langle [A(\tau ),B]_{\eta }\right\rangle _{q},  \label{qSDw}
\end{equation}
in terms of which it is easy to write the spectral representation of the
associated $q$-GF's 
\begin{equation}
G_{qAB}^{(\nu )}\left( \omega \right) =\int_{-\infty }^{+\infty }\frac{%
d\omega ^{\prime }}{2\pi }\frac{\Lambda _{qAB}(\omega ^{\prime })}{\omega
-\omega ^{\prime }+(-1)^{\nu }i\varepsilon },\text{ \ \ }\varepsilon
\rightarrow 0^{+},  \label{eq:GqnSR}
\end{equation}
where $(-1)^{\nu }$ stands for $+1$ if $\nu =r$ and $-1$ if $\nu =a$. These
functions can be analitically continued in the complex $\omega $-plane and
combined to construct the single $q$-GF $G_{qAB}\left( \omega \right)
=\int_{-\infty }^{+\infty }\frac{d\omega ^{\prime }}{2\pi }\frac{\Lambda
_{qAB}(\omega ^{\prime })}{\omega -\omega ^{\prime }}$ of complex $\omega $
with a cut along the real axis.

As a next step, one can immediately obtain the spectral decomposition of the 
$q$-SD. Indeed, by using Eqs. (\ref{Pq(Ei)})-(\ref{Uq(Ei)}), from the
definition (\ref{qSDw}) we have the exact representation: 
\begin{equation}
\Lambda _{qAB}(\omega )=\frac{2\pi }{\overline{Z}_{q}}\sum_{nm}p_{n}^{q}%
\left[ 1+\eta \left( \frac{p_{m}}{p_{n}}\right) ^{q}\right]
A_{nm}B_{mn}\delta \left( \omega -\omega _{mn}\right) ,  \label{SDsp}
\end{equation}
with $\omega _{mn}=\varepsilon _{m}-\varepsilon _{n}$. Then, the spectral
decomposition for $G_{qAB}^{(\nu )}\left( \omega \right) $ follows directly
from the $\omega $-representation (\ref{eq:GqnSR}). In particular, for $%
G_{qAB}\left( \omega \right) $ we have 
\begin{equation}
G_{qAB}(\omega )=\frac{1}{\overline{Z}_{q}}\sum_{nm}\left[ p_{n}^{q}+\eta
p_{m}^{q}\right] \frac{A_{nm}B_{mn}}{\omega -\omega _{mn}}.
\label{SpcDcm_Gqt}
\end{equation}
In strict analogy with the extensive case \cite
{Kalashnikov73,Campana85,CSDM06}, Eq. (\ref{SpcDcm_Gqt}) suggests that the
real poles $\omega _{mn}$ of $G_{qAB}(\omega )$ represent the exact energy
spectrum of undamped excitations in the system.

A comparison of the previous relations with the corresponding extensive ones 
\cite{Kalashnikov73,Campana85} suggests that the Tsallis' statistics does
not influence the meaning of GF singularities, but drastically modifies the
structure of the spectral weights introducing a mixing of energy levels.

\section{The $q$-spectral density method ($q$-SDM)}

Now we have all the necessary ingredients to extend the SDM in the Tsallis'
formalism. By successive derivatives of $\Lambda _{qAB}(\tau )=\left\langle
[A(\tau ),B]_{\eta }\right\rangle _{q}=\int_{-\infty }^{+\infty }d\omega
\Lambda _{qAB}(\omega )e^{-i\omega \tau }/2\pi $ with respect to $\tau $ and
than setting $\tau =0$, one obtains the infinite set of exact equations of
motion or \textit{sum rules} for $\Lambda _{qAB}(\omega )$

\begin{equation}
\int_{-\infty }^{+\infty }\frac{d\omega }{2\pi }\omega ^{m}\Lambda
_{qAB}\left( \omega \right) =\left\langle [L_{H}^{m}A,B]_{\eta
}\right\rangle _{q},\text{ \ }(m=0,1,2,...),  \label{eq:SDM}
\end{equation}
where the quantity on the left-hand side of Eq. (\ref{eq:SDM}) is called the 
$\mathit{m}$\textit{-moment} of $\Lambda _{qAB}(\omega )$ and $L_{H}^{m}$ ($%
m=0,1,2,...$) acts as $L_{H}^{0}A=A$, $L_{H}^{1}A=\left[ A,H\right] _{-}$, $%
L_{H}^{2}A=\left[ \left[ A,H\right] _{-},H\right] _{-}$ and so on.

The set of integral equations (\ref{eq:SDM}) represents a typical \textit{%
moment problem} which should determine exactly the unknown $q$-SD. Of
course, in practical calculations we must look for an approximated solution
of $\Lambda _{qAB}\left( \omega \right) $, which captures the essential
physics of the system under study, truncating the set (\ref{eq:SDM}) at a
given order.

Suggested by the exact spectral decomposition (\ref{SDsp}), as a suitable
approximation for $\Lambda _{qAB}(\omega )$, one can assume a finite sum of
properly weighted $\delta $-functions (the so called \textit{polar ansatz}) 
\begin{equation}
\Lambda _{qAB}(\omega )=2\pi \sum\limits_{k=1}^{n}\lambda _{qAB}^{(k)}\delta
(\omega -\omega _{qAB}^{(k)}),  \label{eq:PolarAnsatz}
\end{equation}
where $n$ is a finite integer number and the unknown parameters $\lambda
_{qAB}^{(k)}$ and $\omega _{qAB}^{(k)}$ are to be determined
self-consistently solving the first $2n$ moment equations in the set (\ref
{eq:SDM}).

In several situations of experimental interest, also the damping of
excitations may be quite relevant so that the polar approximation (\ref
{eq:PolarAnsatz}) is inadequate. In this case, to take properly into account
the finite life-time of quasi-particles, one can assume, as in the extensive
case \cite{Campana85,CSDM06}, the so called \textit{modified Gaussian ansatz}
\begin{equation}
\Lambda _{qAB}(\omega )=2\pi \omega \sum\limits_{k=1}^{n}\lambda _{qAB}^{(k)}%
\frac{\exp \left[ -\left( \omega -\omega _{qAB}^{(k)}\right) ^{2}/\Gamma
_{qAB}^{(k)}\right] }{\sqrt{\pi \Gamma _{qAB}^{(k)}}},
\end{equation}
where $\Gamma _{qAB}^{(k)}$ represents the width of the peak at $\omega
=\omega _{qAB}^{(k)}$ and the life-time of the excitations with frequency $%
\omega _{qAB}^{(k)}$ is given by $\tau _{qAB}^{(k)}=\sqrt{\Gamma _{qAB}^{(k)}%
}$ under the condition $\Gamma _{qAB}^{(k)}{/}\left[ \omega _{qAB}^{(k)}%
\right] ^{2}\ll 1$. It is worth noting that, the $q$-SDM is quite general
and can be easily applied to different models by introducing the appropriate 
$q$-SD.

\section{The quantum spin-$1/2$ Heisenberg ferromagnet with long-range
interactions: a $q$-SDM approach to the lowest order of approximation}

The quantum $d$-dimensional spin-$1/2$ Heisenberg ferromagnet with
long-range interactions \cite{Nakano95} is described in the $\mathbf{k}$%
-space by the Hamiltonian 
\begin{equation}
H=-\frac{1}{2N}\sum_{\mathbf{k}}J\left( \mathbf{k}\right) \left( S_{\mathbf{k%
}}^{+}S_{-\mathbf{k}}^{-}+S_{\mathbf{k}}^{z}S_{-\mathbf{k}}^{z}\right)
-hS_{0}^{z}.  \label{eq:Hmodel}
\end{equation}
Here $h$ is the external magnetic field, $N$ is the number of sites $\left\{
j\right\} $ of a hypercubic lattice with unitary spacing; $\mathbf{S}_{%
\mathbf{k}}$, $S_{\mathbf{k}}^{\pm }$ and $J\left( \mathbf{k}\right) $ are
the Fourier trasforms of the spin operators $\mathbf{S}_{j}$, $S_{j}^{\pm
}=S_{j}^{x}\pm iS_{j}^{y}$ and the exchange interaction $J_{ij}=J/\left| 
\mathbf{r}_{i}\mathbf{-r}_{j}\right| ^{p}$ ($p>0$, $J>0$), respectively, and
the $\mathbf{k}$-sum is restricted over the first Brillouin zone ($1BZ$).

For spin model (\ref{eq:Hmodel}), the $q$-expectation value of the
magnetization per site is given by 
\begin{equation}
m_{q}=\frac{1}{N}\sum_{j=1}^{N}\left\langle S_{j}^{z}\right\rangle _{q}=%
\frac{1}{2}-\frac{1}{N^{2}}\sum_{\mathbf{k}}\left\langle S_{-\mathbf{k}%
}^{-}S_{\mathbf{k}}^{+}\right\rangle _{q}.  \label{eq:Mq}
\end{equation}
Due to the operatorial representation (\ref{eq:Hmodel})-(\ref{eq:Mq}), in
the framework of the $q$-SDM it is convenient to introduce the $q$-SD 
\begin{equation}
\Lambda _{q\mathbf{k}}(\omega )=\int_{-\infty }^{\infty }dte^{i\omega \tau
}\left\langle \left[ S_{\mathbf{k}}^{+}\left( \tau \right) ,S_{-\mathbf{k}%
}^{-}\right] _{-}\right\rangle _{q},
\end{equation}
which satisfy the infinite hierarchy of moment equations 
\begin{equation}
\int_{-\infty }^{+\infty }\frac{d\omega }{2\pi }\omega ^{m}\Lambda _{q%
\mathbf{k}}(\omega )=\left\langle [L_{H}^{m}S_{\mathbf{k}}^{+},S_{-\mathbf{k}%
}^{-}]_{-}\right\rangle _{q},\text{ }(m=0,1,2,...).\text{\ }
\label{eq:set18}
\end{equation}
To the lowest order of approximation, we can assume for $\Lambda _{q\mathbf{k%
}}(\omega )$ the one $\delta $-function polar ansatz 
\begin{equation}
\Lambda _{q\mathbf{k}}(\omega )=2\pi \lambda _{q\mathbf{k}}\delta (\omega
-\omega _{q\mathbf{k}})\text{,}  \label{eq:SD_delta}
\end{equation}
where the unknown parameters $\lambda _{q\mathbf{k}}$ and $\omega _{q\mathbf{%
k}}$ can be determined by solving the first $2$ moment equations of the set (%
\ref{eq:set18}). Then, working close to $q=1$ and with the near saturation
condition $\left\langle S_{\mathbf{k}}^{z}S_{-\mathbf{k}}^{z}\right\rangle
\approx \left\langle S_{\mathbf{k}}^{z}\right\rangle \left\langle S_{-%
\mathbf{k}}^{z}\right\rangle =N^{2}m_{q}^{2}\delta _{\mathbf{k},0}$ \cite
{CSDM04}, we obtain for $\omega _{q\mathbf{k}}$ and $m_{q}$ the set of
self-consistent equations (with $\lambda _{q\mathbf{k}}=2Nm_{q}$) 
\begin{eqnarray}
\omega _{q\mathbf{k}} &=&h+J\Omega _{p}(\mathbf{k})m_{q}+\frac{J}{N}\sum_{%
\mathbf{k}^{\prime }}\frac{\left[ \Omega _{p}(\mathbf{k}-\mathbf{k}^{\prime
})-\Omega _{p}(\mathbf{k}^{\prime })\right] }{1-\left[ 1-\beta \left(
1-q\right) \omega _{q\mathbf{k}^{\prime }}\right] ^{\frac{q}{1-q}}},
\label{eq:Eq1} \\
m_{q} &=&\frac{1}{2}-\frac{1}{N}\sum_{\mathbf{k}}\frac{2m_{q}}{\left[
1-\beta \left( 1-q\right) \omega _{q\mathbf{k}}\right] ^{\frac{q}{q-1}}-1},
\label{eq:Eq2}
\end{eqnarray}
where 
\begin{equation}
\Omega _{p}(\mathbf{k})=\frac{J(0)-J(\mathbf{k})}{J}=\sum_{\mathbf{r}}\frac{%
1-\cos \mathbf{k}\cdot \mathbf{r}}{|\mathbf{r}|^{p}}.  \label{eq:Omega(k)}
\end{equation}
It is worth noting that in Eq. (\ref{eq:Eq1}), which determines the $q$%
-excitation spectrum $\omega _{q\mathbf{k}}$, the contribution $h+J\Omega
_{p}(\mathbf{k})m_{q}$ is formally identical to the known Tyablikov
dispersion relation \cite{GFbooks,Tyablikov67} for the corresponding
extensive problem.

The solution of the self-consistent problem (\ref{eq:Eq1})-(\ref{eq:Omega(k)}%
) is rather complicate and one must consider asymptotic regimes for
obtaining explicit analytical results. For instance, in the low-temperature
ferromagnetic phase, we can resort to the Tyablikov-like approximation 
\begin{equation}
\omega _{q\mathbf{k}}\approx \omega _{\mathbf{k}}^{(T)}=h+\frac{1}{2}J\Omega
_{p}(\mathbf{k}),  \label{eq:Mq-Tyab}
\end{equation}
as a zero order in Eqs. (\ref{eq:Eq1})-(\ref{eq:Eq2}). So, in the
thermodynamic limit $N\rightarrow \infty $, Eq. (\ref{eq:Eq2}) yelds a
cumbersome expression for the $q$-magnetization $m_{q}(\beta ,h)$ which, for 
$q=1$, reproduces the extensive counterpart \cite{Nakano95} and can be used
to obtain explicit expansion as $q\rightarrow 1$.

Interesting representation for $m_{q}(\beta ,h)$ and the $q$-susceptibility $%
\chi _{q}\left( \beta ,h\right) =\partial m_{q}(\beta ,h)/\partial h$ in the
nearly saturation regime can be obtained, for $d<p<2d$, under condition $%
\beta h(q-1)>1$. Assuming the low-$k$ expansion $\Omega _{p}(\mathbf{k}%
)\approx A_{d}(p)k^{p-d}$ \cite{Nakano95}, where $A_{d}(p)=\pi ^{d}d^{d-p}%
\left[ \Gamma (p)\right] ^{-d}/\sin [\pi (p-d)/2]$ and $\Gamma (z)$ the
gamma function, we find indeed 
\begin{eqnarray}
m_{q}(\beta ,h) &\simeq& \frac{1}{2}-\frac{K_{d}\Lambda ^{d}}{d\left( 1+\beta
h(q-1)\right) ^{\frac{q}{q-1}}} \text{ }_{2}F_{1}\left( \frac{d}{p-d},\frac{q}{q-1};%
\frac{p}{p-d};-\frac{JA_{d}(p)\beta (q-1)\Lambda ^{p-d}}{2\left( 1+\beta
h(q-1)\right) }\right),  \notag \\
\label{eq:24} \\
\chi _{q}\left( \beta ,h\right) &\simeq& \frac{K_{d}\Lambda ^{d}\beta }{%
(p-d)\left( 1+\beta h(q-1)\right) ^{\frac{2q-1}{q-1}}}\left\{ \frac{q-1}{%
\left[ 1+\frac{JA_{d}(p)\beta (q-1)\Lambda ^{p-d}}{2\left( 1+\beta
h(q-1)\right) }\right] ^{\frac{q}{q-1}}}\right.   \notag \\
&& \left. + \left[ 1-(2-\frac{p}{d})q\right] \text{ }_{2}F_{1}\left( \frac{d}{p-d%
},\frac{q}{q-1};\frac{p}{p-d};-\frac{JA_{d}(p)\beta (q-1)\Lambda ^{p-d}}{%
2\left( 1+\beta h(q-1)\right) }\right) \right\} .  \label{eq:25}
\end{eqnarray}
Here, $_{2}F_{1}\left( a,b;c;z\right) $ is the hypergeometric function, $%
K_{d}=2^{1-d}\pi ^{-d/2}/\Gamma (d/2)$ and $\Lambda $ is a wave vector
cut-off related to the $1BZ$ of the spin lattice.

If, additionally, we assume $\beta h(q-1)\gg 1$ as $T=\beta ^{-1}\rightarrow
0$, Eqs. (\ref{eq:24}) and (\ref{eq:25}) yeld $m_{q}(\beta ,h)\simeq 1/2-%
\mathcal{A}_{d}^{(q)}(h)T^{\frac{q}{q-1}}$ and $\chi _{q}\left( \beta
,h\right) \simeq \mathcal{B}_{d}^{(q)}(h)T^{\frac{q}{q-1}}$, where the
explicit expressions of $\mathcal{A}_{d}^{(q)}(h)$ and $\mathcal{B}%
_{d}^{(q)}(h)$\ (with $h\neq 0$) are inessential for our purposes.

The intrinsically nonextensive region ($p\leq d$), which is not of primary
interest in this short contribution, requires a more delicate analysis which
will be the subject of a future work.

\section{Concluding remarks}

In this short note we have extended, in nonextensive quantum
thermostatistics, the two-time GF formalism and the related SDM already
developed for quantum \cite{Kalashnikov73,Campana85} and classical \cite
{Campana84,CSDM04,CSDM06} extensive systems. This offers the possibility to
explore, at least in principle, the properties of realistic systems by using
the big amount of experiences acquired in extensive problems. In case of the
Heisenberg model (\ref{eq:Hmodel}), the polar ansatz (\ref{eq:SD_delta})
yields reasonable results for the excitation spectrum and the relevant
thermodynamic $q$-quantities in the low temperature regime. For describing
other thermodynamic regimes in a wider range of temperatures, one could to
adopt a new set of approximations involving additional decoupling procedures
and higher order moment equations, consistently with the spirit of the SDM 
\cite{Kalashnikov73,Campana84,Campana85,CSDM04,CSDM06}.


\end{document}